\documentclass[10pt,draft]{dis03}
\usepackage{epsf,amsmath}

\begin{document}
\title{NuTeV Cross Section and \\Structure
Function Measurements}

\author{D. Naples$^7$, T. Adams$^4$, A.~Alton$^4$,
  S.~Avvakumov$^8$, L.~de~Barbaro$^5$,\\
 P.~de~Barbaro$^8$,
  R.~H.~Bernstein$^3$, A.~Bodek$^8$, T.~Bolton$^4$, S.~Boyd$^7$, \\
  J.~Brau$^6$,
 D.~Buchholz$^5$, H.~Budd$^8$, L.~Bugel$^3$,
  J.~Conrad$^1$, 
R.~B.~Drucker$^6$, \\
 B.~T. Fleming$^1$, J.~Formaggio$^1$, R.~Frey$^6$,
  J.~Goldman$^4$, M.~Goncharov$^4$, \\
D.~A.~Harris$^8$, J.~H.~Kim$^1$,
  S.~Koutsoliotas$^1$, R.~A.~Johnson$^2$, M.~Lamm$^3$,  \\
W.~Marsh$^3$, D.~Mason$^6$, J.~McDonald$^7$,
K.~McFarland$^8$, C.~McNulty$^1$,\\
   P.~Nienaber$^3$, V.~Radescu$^7$, A.~Romosan$^1$,
  W.~Sakumoto$^8$, H.~Schellman$^5$, \\
 M.~H.~Shaevitz$^1$, P.~Spentzouris$^1$, E.~G.~Stern$^1$, N.~Suwonjandee$^2$,\\
  N.~Tobien$^3$, 
M.~Tzanov$^7$, A.~Vaitaitis$^1$, M.~Vakili$^2$,\\
  U.~K.~Yang$^8$, J.~Yu$^3$, G.~P.~Zeller$^5$,
  E.~D.~Zimmerman$^1$}
\maketitle
\vspace{0.1in}

\begin{center}
\it {NuTeV Collaboration\\
 $^1$Columbia University, New York,
  NY, $^2$University of Cincinnati, Cincinnati, OH, $^3$Fermi National
  Accelerator Laboratory, Batavia, IL, $^4$Kansas State University,
  Manhattan, KS, $^5$Northwestern University, Evanston, IL,
  $^6$University of Oregon, Eugene, OR, $^7$University of Pittsburgh,
  Pittsburgh, PA, $^8$University of Rochester, Rochester, NY.}

\end{center}

\begin{abstract}

The NuTeV experiment has obtained a unique
high statistics sample of neutrino and antineutrino
interactions using its high-energy sign-selected
beam. 
Charged-current 
$\nu$ and $\overline{\nu}$ 
differential 
cross sections are extracted. 
Neutrino-Iron structure functions, $F_2(x,Q^2)$ and $xF_3(x,Q^2)$,
are determined by fitting the $y$-dependence of the 
differential cross sections.
NuTeV has precise understanding
of its hadron and muon energy scales, which improves
the systematic precision of this measurement. 
\end{abstract}

\section{Introduction} 

Neutrino deep-inelastic scattering offers a precise
and complimentary probe of nucleon structure and QCD.
Neutrinos only experience the weak force and are
uniquely sensitive to the parity violating structure function,
$xF_3(x,Q^2)$. 

NuTeV is a second generation neutrino deep inelastic
scattering experiment which
used the refurbished CCFR detector and Fermilab's Sign-Selected Quadrapole
Train (SSQT) beamline to produce
seperate high purity neutrino
and antineutrino beams.
QCD results benefit from the precise determination of
muon and hadron energy scales,
and measured detector response functions obtained using NuTeV's
precision calibration beam.
Muon energy scale was determined to 0.7\% and
hadron scale to 0.43\%. 
A more detailed description of the experiment and the
precision calibration can be found in \cite{nutcal}.

\section{Cross Section Measurement} 

The differential cross section in
$x$, the Bjorken scaling variable, and
$y$, the inelasticity, 
is determined from, 
\begin{equation}
  \frac{d^2\sigma^{\nu(\overline{\nu})}(E)}{dxdy} =  \frac{1}{\Phi(E)}\frac{d^2 N^{\nu
(\overline{\nu})}(E)}{dxdy}. 
\label{eq:dfe1}
\end{equation}
where $\Phi(E)$ is the relative flux at neutrino energy $E$.
Beam energy ranges from 30-350~GeV.
Data selection for this sample requires event containment, 
a momentum analyzed muon in the toroid spectrometer, and
minimum energy requirements; hadronic energy, $E_{HAD}>10$~GeV,
muon energy, $E_{\mu}>15$~GeV, and reconstructed
neutrino energy $E_{\mu}>30$~GeV. 
$Q^2>1$ is required to minimize the non-perturbative contributions
in the cross section.
A total of $8.6\times10^5$ neutrino and
$2.3\times10^5$ anti-neutrino
events passed selection requirements.

Neutrino flux is determined from a sample of 
events at low $E_{HAD}<20$~GeV using the ``fixed $\nu_o$'' 
method \cite{tim}. The integrated number of events with 
$E_{HAD}<20~GeV$ as $y=\frac{E_{HAD}}{E_{\nu}}\rightarrow 0$
is proportional to the flux. Corrections, determined from the
data,  up to order $y^2$ are applied to determine the relative 
flux to about the 1\% level. Flux is normalized using the world
average $\nu-Fe$ cross section $\frac{\sigma^\nu}{E}=0.677\times 10^{-38}$
cm$^2$/GeV from reference \cite{pdg}. 

The Monte Carlo simulation , used to account for acceptance and 
resolution effects, requires an input cross section model
which is iteratively tuned to fit the data. 
Cross section data are fit to empirically determine a set of parton
distribution functions
and the procedure is iterated 
until convergence.

\begin{figure}[!thb]
\vspace*{7.2cm}
\includegraphics{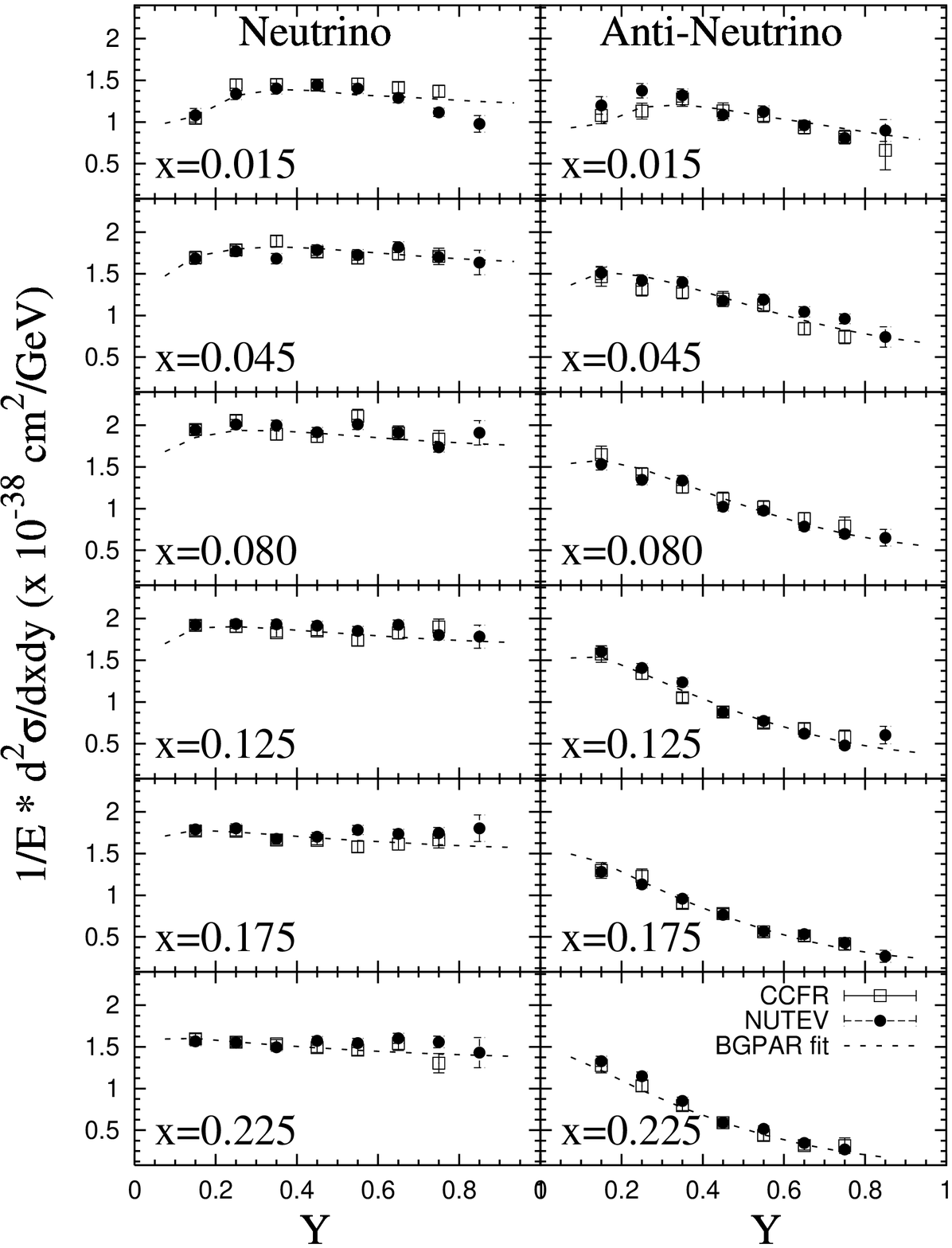}
\includegraphics{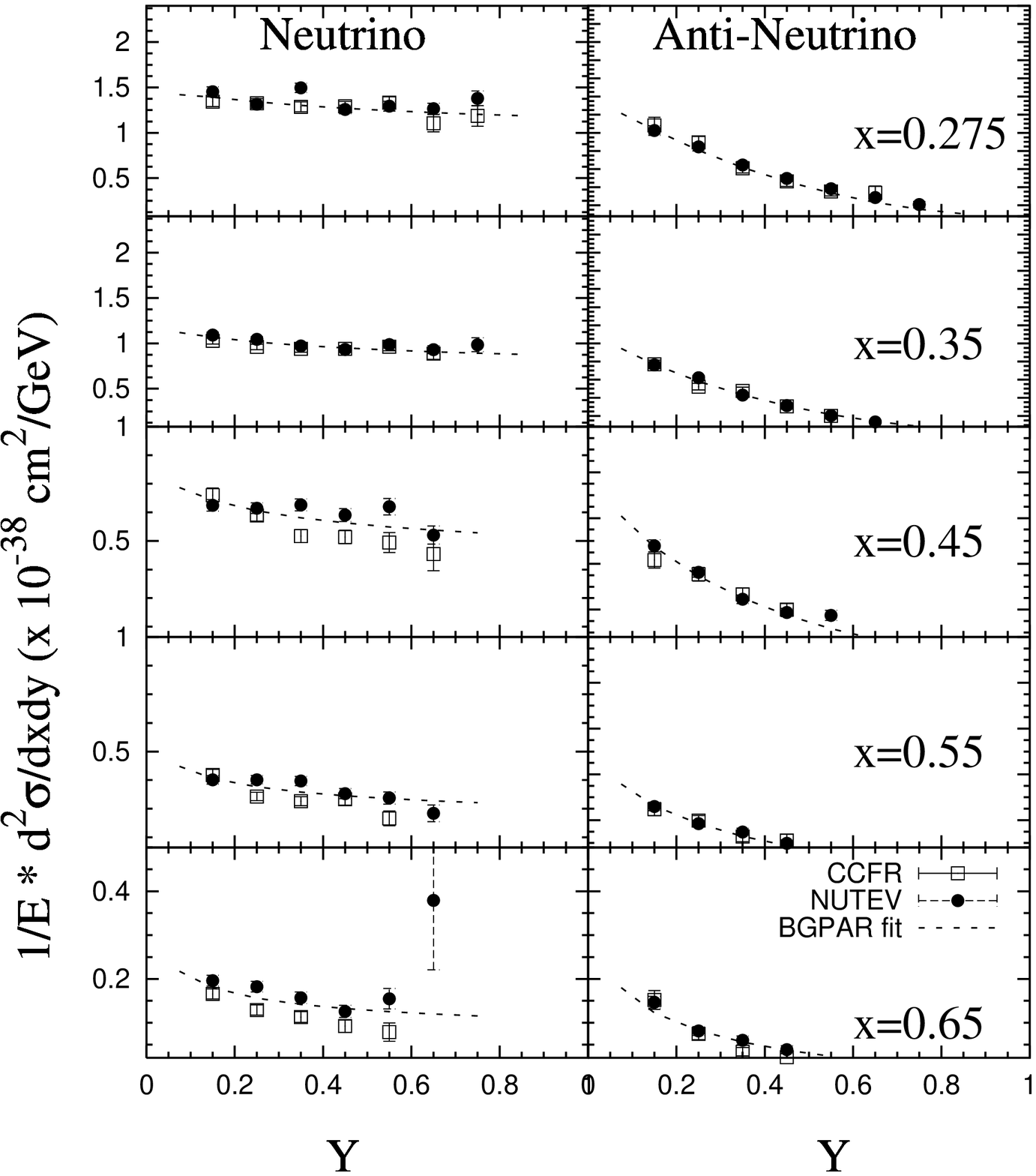}
\caption{Preliminary differential cross sections for
neutrinos (left each side) and anti-neutrinos (right each side) at $E=85$ GeV. 
Points are NuTeV (filled circles) and 
CCFR (open squares). Curve shows fit to NuTeV data.}
\label{diffxsec}
\end{figure}

Figure \ref{diffxsec} shows the preliminary extracted
cross sections
compared with the CCFR \cite{ukthesis}
for $E_{\nu}=85~GeV$. 
NuTeV's $\overline{\nu}$ cross section 
is in good agreement at all $x$ with CCFR. 
The neutrino cross section agrees for 
$x<0.45$ but is systematically above CCFR for 
high-$x$ at high $y$. In the 
high-$x$, high-$y$ region, smearing effects are larger
and detector modeling is more important.

\section{Structure Functions}

The sum of the differential
cross sections can be expressed as:
\begin{equation}
      \left[ \frac{d^2\sigma^\nu}{dxdy} +
    \frac{d^2\sigma^{\overline{\nu}}}{dxdy} \right]
= 
\frac{y^2 G_F^2 M E}{\pi (1-\epsilon)}
\left[\left(1+\epsilon R_L\right)2xF_1 + 
\frac{y(1-y/2)}{1+(1-y)^2}\Delta xF_3 \right]
\end{equation}
where $G_F$ is the Fermi weak coupling constant and M is the 
proton mass.
The $y$-dependence of the first term is implicit in
$\epsilon =  \frac{2(1-y)- Mxy/E}{1+(1-y)^2 + Mxy/E}$,
the polarization of the virtual $W$-boson.
$R_L(x,Q^2)$ relates the structure functions $F_2(x,Q^2)$ 
and $2xF_1(x,Q^2)$ by,
$F_2(x,Q^2) \approx 2xF_1{(1 + R_L(x,Q^2))}$
In the last term, $\Delta xF_3=xF_3^{\nu}-xF_3^{\overline{\nu}}$, is sensitive
to the heavy-quark seas, $\Delta xF_3\approx4x\left(s-c\right)$.

To extract $F_2(x,Q^2)$ we use
$\Delta xF_3$ from a NLO QCD model as input (TRVFS \cite{trvfs}) 
and $R_L(x,Q^2)$ from $R_L^{\scriptscriptstyle{world}}$ \cite{rworld}.
Cross sections are corrected for small excess of neutrons
in the target (5.67\%) and for radiative effects \cite{bardin}
before fits are performed. 
NuTeV's $F_2(x,Q^2)$ is shown in Figure 2
compared
with CCFR\cite{ukthesis}.
$F_2(x,Q^2)$ is in good agreement with CCFR for $x<0.45$
and systematically above at higher $x$
consistent with the higher measured $\nu$ cross section in this region.
The curve shown is a NLO QCD model from \cite{trvfs}.

The difference of 
differential cross
sections is proportional to $xF_3(x,Q^2)$,
\begin{equation}
\left[\frac{d^2\sigma^{\nu}}{dxdy} -
      \frac{d^2\sigma^{\overline{\nu}}}{dxdy}\right]=
\frac{2 G_F^2 M E}{\pi}
\left(y-\frac{y^2}{2}\right) xF_3^{\scriptstyle AVG}(x,Q^2)
\end{equation}
where $xF_3^{\scriptstyle AVG}=\frac{1}{2}(xF_3^{\nu}+xF_3^{\overline{\nu}})$.
Figure 2
shows the NuTeV measurement of $xF_3(x,Q^2)$ 
which is in agreement with CCFR \cite{seligman}
for $x<0.45$ and again systematically above for higher $x$.

\begin{figure}[!thb]
\vspace*{7.0cm}
\begin{center}
\includegraphics{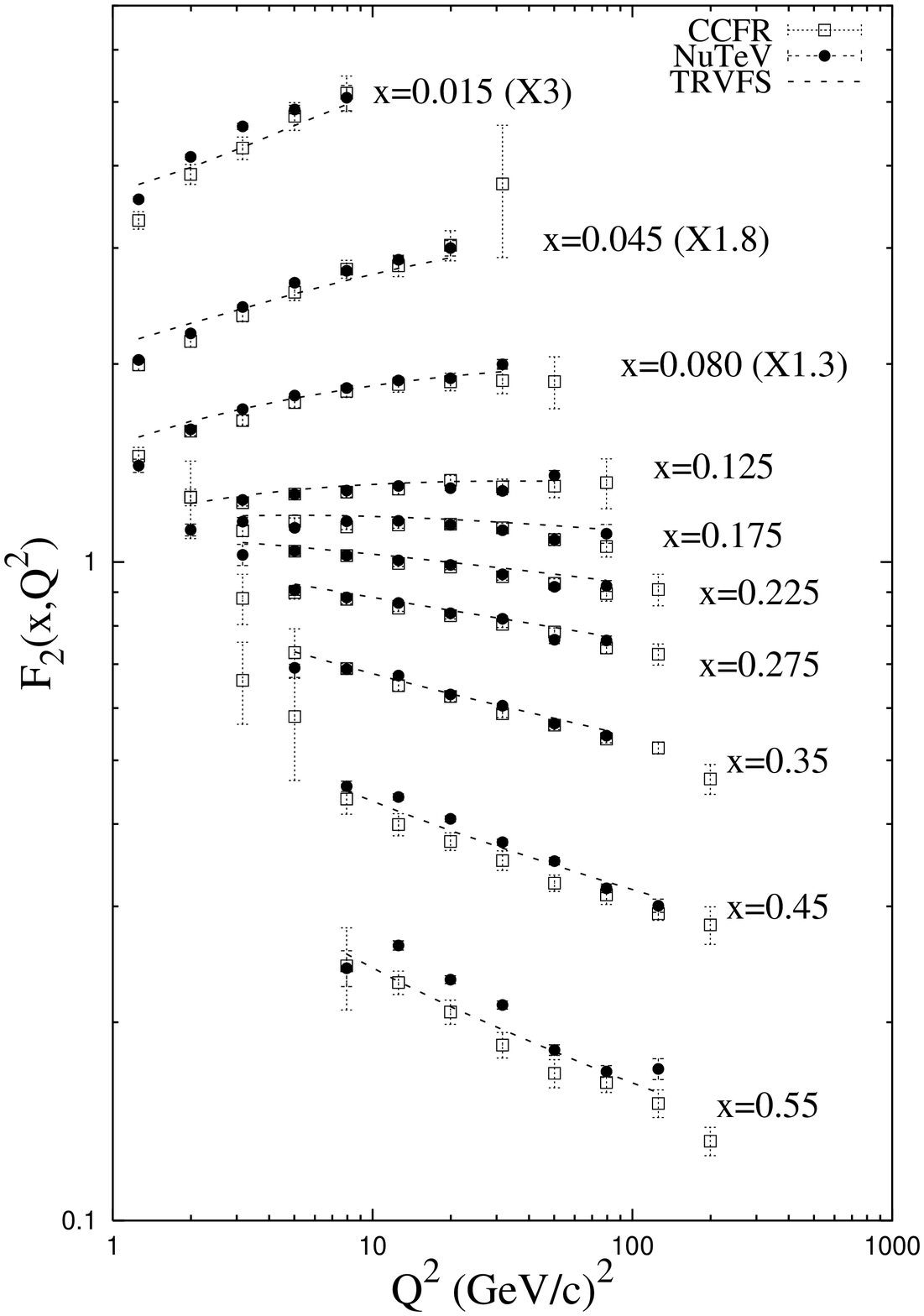}
\includegraphics{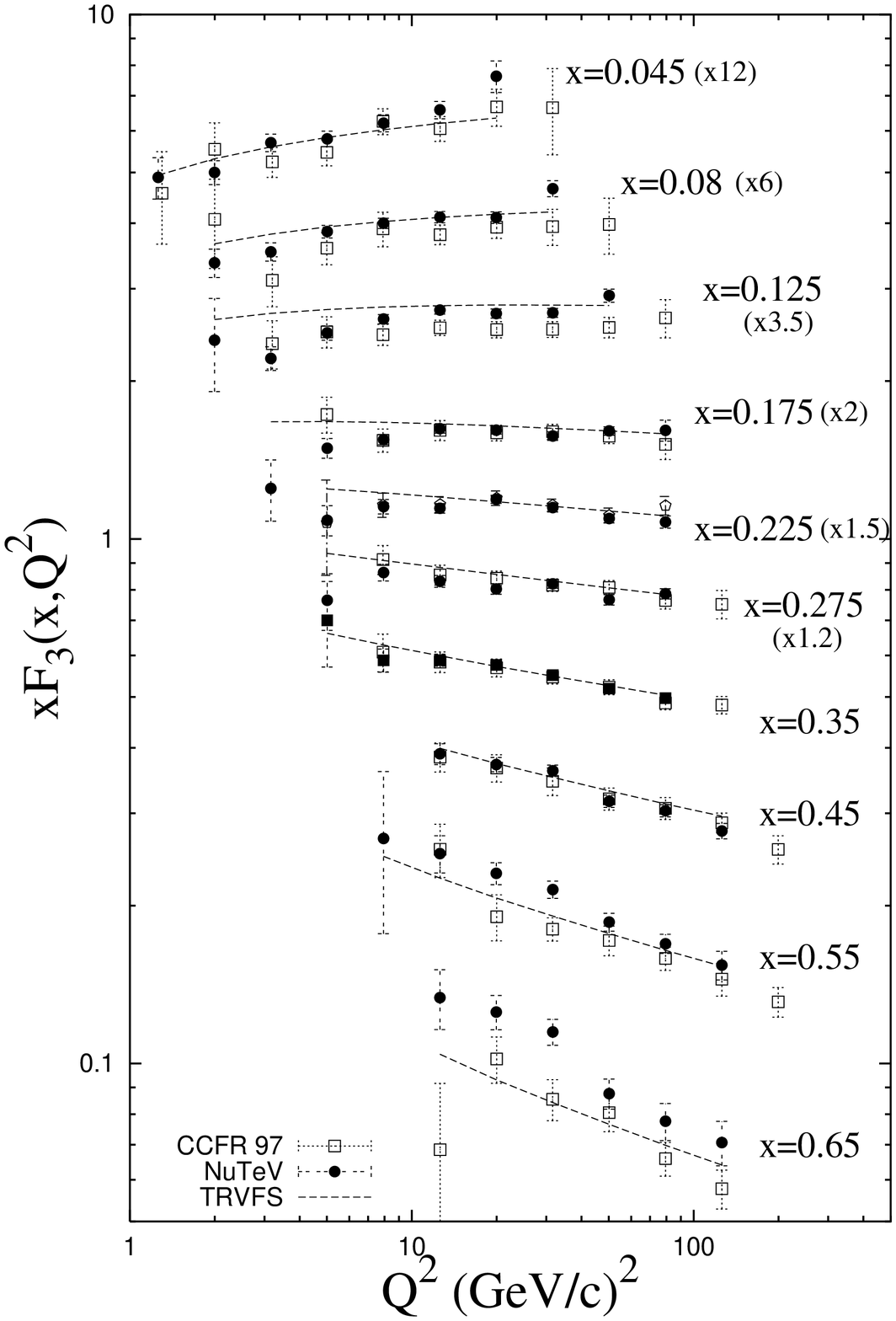}
\caption{(Left) NuTeV's Preliminary Measurement of $F_2(x,Q^2)$
compared with previous $\nu$-Fe results from CCFR \cite{ukthesis}. 
(Right) $xF_3(x,Q^2)$ extracted from 
the cross section difference and a comparison with the measurement
from CCFR \cite{seligman}.
Curves show Thorne-Roberts NLO model \cite{trvfs}.}
\end{center}
\label{f2}
\end{figure}

NuTeV plans to extend the 
differential cross section sample 
to a previously inaccessible high-$y$ range.
The sign-selected beam allows
NuTeV to use a low energy muon sample, (with $E_\mu$ down to $4$~GeV),
with momentum determined
from range in the calorimeter and
muon sign assumed by
the running mode ({\em i.e.} $\nu$ or $\overline{\nu}$).
The addition of this sample will provide a lever
arm to better constrain $R_L$ in two-parameter fits.

\section{Conclusions}

NuTeV has extracted preliminary differential cross sections for 
$\nu$-Fe and $\overline{\nu}$-Fe deep-inelastic scattering. 
Differential cross sections provide the most model-independent observable 
which can be reported for this process.
NuTeV has precise understanding of muon and hadron 
energy scales, and a detector response determined over
the entire energy range allowing improved systematic precision
for this measurement.

Preliminary $\nu$-Fe structure functions have been extracted 
from fits to the differential cross section.
QCD fits to the non-singlet structure function 
will provide stringent constraints on $\Lambda_{\scriptstyle QCD}$.

\end{document}